# Dual-band Photonic Filters with Wide Tunable Range Using Chirped Sampled Gratings

Simeng Zhu,* Bocheng Yuan, Weiqing Cheng, Yizhe Fan, Yiming Sun, Mohanad Al-Rubaiee, Jehan Akbar, John H. Marsh, Lianping Hou

*James Watt School of Engineering, University of Glasgow, Glasgow, G12 8QQ, U.K.*
*Corresponding author: 2635935z@student.gla.ac.uk*

**We have developed a photonic filter featuring dual independently tunable passbands. Employing the reconstruction equivalent-chirp technique, we designed linearly chirped sampled Bragg gratings with two equivalent phase shifts positioned at 1/3 and 2/3 of the cavity, thus introducing two passbands in the +1st channel. Leveraging the significant thermo-optic effect of silicon, dual-band tuning is achieved via micro-heaters integrated on the chip surface. By tuning the injection currents ranging from 0 to 35 mA into the micro-heaters, the filter exhibits a wide range of dual-wavelength filtering performance, with the frequency interval between the two passbands adjustable from 37.2 GHz to 186.1 GHz.**

With the ongoing expansion of 5G mobile communications, satellite communications, and Internet of Things (IoT) applications, the demand for multiband communication in wireless Radio Frequency (RF) systems is increasingly growing [1-2]. Photonic filters (PFs), as critical components in microwave photonics, are receiving significant attention due to their outstanding frequency selection, filtration performance, and dynamic tunability [2-4]. Stimulated Brillouin Scattering (SBS)-based PFs have become a popular research topic due to their high resolution and GHz-level tuning capability [5-6]. However, they require centimeter-scale waveguides [6] to generate only a single notch, and the local oscillator (LO) tone required for SBS must be generated through complex external intensity modulators and additional bandpass filters [5]. To meet the dynamic operational demands of multiband communications, the development of adjustable and reconfigurable multiband photonic filters is urgent. Recent studies have reported the design of various tunable dual-band filters, which can be implemented through several approaches. These include two cascaded microring resonators (MRR) with a tuning range of 6-17 GHz [7], electro-optic (EO) modulation-based distributed feedback Bragg-grating resonators (DFBR) with a -3 dB bandwidth of 24 GHz and a cavity length of 875 μm each [8], or using single resonator configurations such as equivalent phase-shift fiber Bragg grating (EPS-FBG) with a tuning range of only 7.4 GHz [9], the dual-arm Mach-Zehnder interferometer (MZI) with a tuning range of 1-18 GHz [10], and dual phase-shifted Bragg grating (PS-BG) with a tuning range of 113.4 GHz, though the passbands cannot be tuned independently [11].

In this paper, we designed and fabricated a dual-bandpass filter based on silicon on insulator (SOI) platform, achieving independent tuning of two passbands on a single grating based on photonic integrated waveguide for the first time. The filtering section of the device consists of an equivalent chirped sampled grating (CSG) and two equivalent π phase shifts (EPS). The two equivalent phase shifts are designed to generate two passbands in the +1st channel. The equivalent chirp of the grating spatially separates the photon distributions of the two passbands, enabling independent tuning of the passband wavelengths. Meanwhile, this equivalent chirp is achieved by linearly modulating the micron-scale sampling period, which significantly reduces manufacturing difficulty compared to modulating the nanometer-scale seed grating period [12]. Two micro-heaters (MH) are respectively placed at the two EPSs locations. By utilizing the thermo-optic (TO) effect to change the phase magnitude of two EPSs, the central wavelength of the passbands can be adjusted.

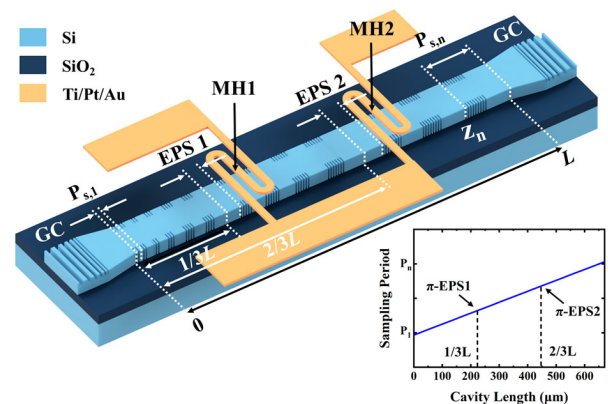

**Fig. 1.** Schematic of the CSG includes two π-equivalent phase shifts (π-EPSs), integrated input and output GCs, and two MHs positioned above the EPSs.

Compared to conventional MZI and MR, the device proposed here offers narrow passband frequencies and a large frequency tunable range unrestricted by the Free Spectral Range (FSR) [3], presenting unique advantages in wavelength selectivity and multiplexing capabilities, which are particularly critical for RF manipulation. Simultaneously, our approach minimizes the quantity of resonators needed to process multiple signals, presenting a compact solution for photonic integrated circuits (PICs).

Fig. 1 is a schematic of the proposed device. The SOI wafer has a 220 nm top silicon layer and a 2 μm buried oxide (BOX) layer on a 675 μm thick silicon substrate. To generate two distinct passbands, two EPSs (EPS1 and EPS2) are strategically embedded at positions $1/3\,L$ and $2/3\,L$ along the CSG cavity's length ($L$).

By calculating the Fourier coefficients of the uniform sampled Bragg gratings, the coupling coefficient $\kappa$ in +1st channel is $1/\pi$ times that of the uniform Bragg grating. Assuming there is a displacement $\Delta L$ inserted at position z of the cavity, then the refractive index modulation $\Delta n(z)$ is

$$\Delta n(z) = \frac{1}{2}\Delta n_0 S_{+1} e^{-i\frac{2\pi z}{\Lambda_0} - i\frac{2\pi z}{P_s} - i\frac{2\pi \Delta L}{P_s}} \tag{1}$$

where $\Delta n_0$ is the refractive index modulation amplitude of the seed grating, $S_{+1}$ represents the Fourier expansion of the +1st channel, $\Lambda_0$ is the seed grating period, and $P_s$ refers to the sampling period. Thus, when $2\pi\Delta L/P_s$ is equal to $\pi$, i.e., $\Delta L$ is equal to $P_s/2$, this corresponds to the introduction of an equivalent π phase shift in the +1st channel [13].

The linear chirp sampling period distribution satisfies the following equation:

$$P_{s,n} = P_{s,1} + C \cdot z_n \tag{2}$$

where $P_{s,1}$ is the first sampling period, $P_{s,n}$ represents the sampling period at position of $z_n$. $C$ refers to the chirp rate, defined as the ratio of the difference between the maximum and minimum sampling periods ($P_{s,n} - P_{s,1}$) to the cavity length $L$. In our device, the chirp rate $C$ is set to nm/mm. By introducing a linear chirp to the sampled grating, the sampling periods at the two EPS locations differ, leading the two passbands to originate from different optical cavities, which enables independent tuning of the passbands.

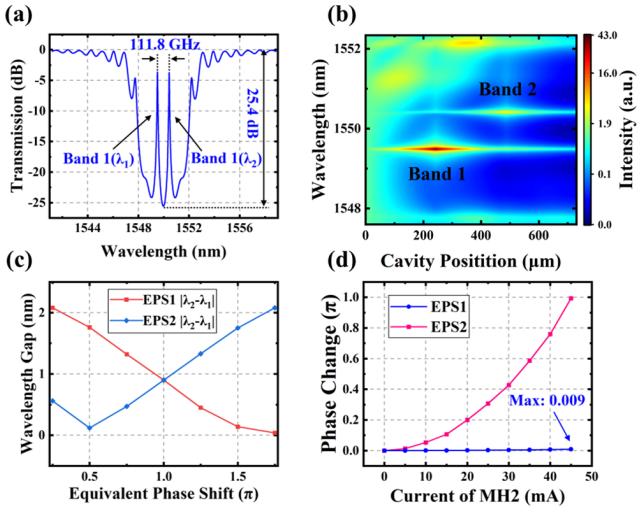

**Fig. 2.** Simulation results of (a) transmission spectrum of the two passbands; (b) photon distribution along the cavity; (c) wavelength drift versus phase shift value of EPS1 and EPS2, and (d) phase change value of EPS1 and EPS2 versus the inject current of MH2

Fig. 2(a) is the transmission spectrum of the proposed device calculated using the transmission matrix method (TMM) [14]. Two narrow passbands (Q factor: $6.2 \times 10^4$), referred to as band 1 and band 2 (respective center wavelengths are $\lambda_1$ and $\lambda_2$) are clearly observed within the transmission stopband with a center wavelength of 1550 nm, aligning with the phase discontinuities caused by EPS1 and EPS2, respectively. The initial frequency spacing between band 1 and band 2 in the filter spectrum is 111.8 GHz, and the extinction ratio (ER) of the grating is 25.4 dB. The spatial photon distribution of the two passbands is shown in Fig. 2(b). After applying the chirp, the photons of $\lambda_1$ and $\lambda_2$ are concentrated at the two EPS locations, spatially separated, which reduces their interaction within the cavity. Therefore, when we modulate the value of a specific EPS, only one corresponding wavelength is adjusted, while the other remains unchanged. When an injection current is applied to the MH, the resistance wire is electrically heated due to the Joule effect, and the resulting heat is conducted downward through the deposited $SiO_2$ cladding (Thermal Conductivity: 1.1 W/mK at room temperature) [15]. Due to the substantial thermo-optic coefficient of silicon ($1.84\times10^{-4}$/K) [16], the refractive index of the buried waveguide changes with temperature, which is equivalent to a change in the phase of the EPS.

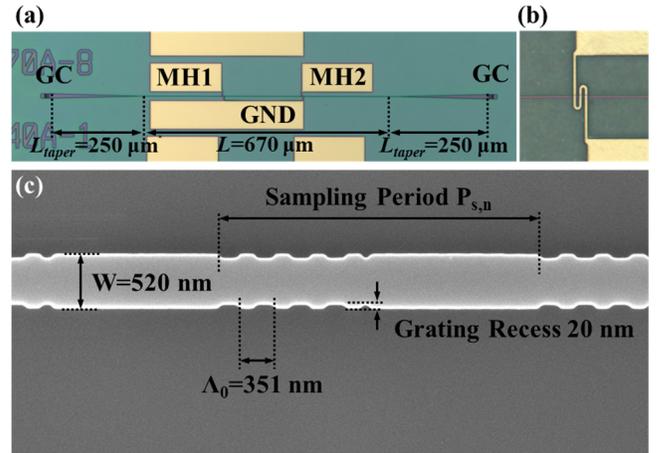

**Fig. 3.** (a) Optical microscope image of the PF device; (b) zoomed-in view of the MH; (c) scanning electron microscope (SEM) image of the CSG.

In Fig. 2(c), we calculated $\lambda_1$ and $\lambda_2$ under different values of EPS1 and EPS respectively while keeping the other EPS constant at π. The results show that increasing the phase shift of EPS1 causes a redshift in $\lambda_1$, while $\lambda_2$ has a slight fluctuation of ± 0.06 nm in the EPS range of 1 π to 1.75 π, resulting in a decrease in wavelength spacing from 111.8 GHz (initially unchanged EPS phase shift at π) to 5 GHz (at 1.75 π). Similarly, increasing the phase shift of EPS2 while keeping EPS1 constant at π results in $\lambda_1$ remaining relatively stable (± 0.02 nm) as EPS2 changes from 1 π to 1.75 π, while $\lambda_2$ shifts away from $\lambda_1$, increasing the wavelength spacing to 259 GHz at 1.75 π. This method achieves modulation of the filter passband spacing. As illustrated in Fig. 2(d), our calculations using COMSOL indicate that the thermal crosstalk is only 0.9% when a 45 mA drive current is applied exclusively to MH2. This indicates that the MH distribution in this device provides excellent thermal isolation, with the thermal effects on adjacent EPSs beyond 225 μm being negligible.

Fig. 3(a) is the optical microscope image of the fabricated device. Where the CSG section ($L$=670 μm) is at the center of the device and the two tapers ($L_{taper}$=250 μm) for connecting the CSG to the grating couplers (GC) are symmetrically integrated on both sides.

Highly localized MHs fabricated on the top cladding of the EPS positions enable independent tuning of EPSs by current injection. The MH consists of contact pads and heating resistance wires, with serpentine heating wires positioned directly above the two EPSs regions, each with a width of 0.9 μm, a spacing of 0.8 μm, and a resistance of 10.7 Ω at 20°C as shown in Fig. 3(b). The MH spans a total waveguide length of 5.4 μm and experiences an additional mode transmission loss of 3.5 dB/cm when placed over the cladding. The extra transmission loss attributed to the two MHs is $6.6 \times 10^{-3}$ dB. As depicted in Fig. 3(c), the side-wall gratings of the CSG are symmetrically arranged, featuring a uniform Bragg grating as the seed grating. This grating is defined by a period ($\Lambda_0$) of 351 nm, a grating recess of 20 nm, and a ridge waveguide width ($W$) of 520 nm. The CSG waveguide has an effective refractive index of 2.47 at a wavelength of 1550 nm.

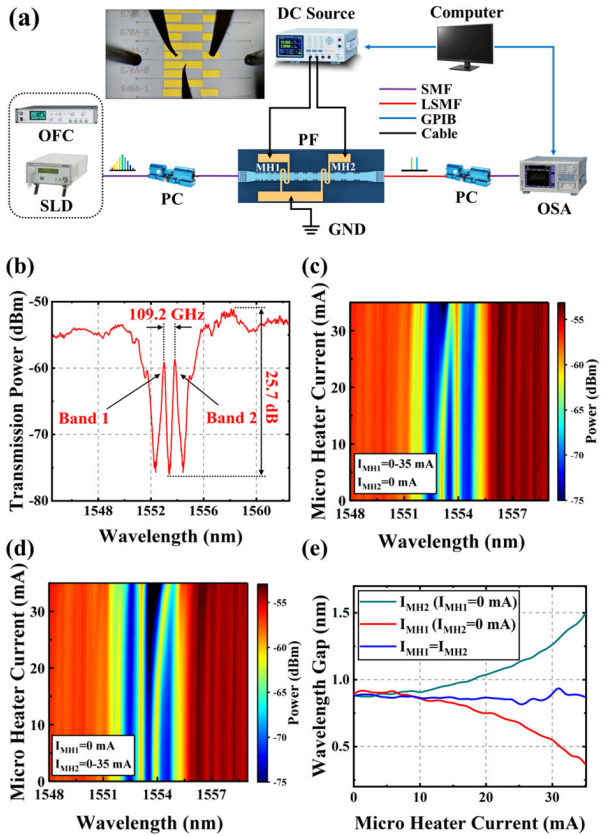

**Fig. 4.** (a) Schematic of the experiment setup; (b) transmission spectrum measured with no current applied to the MHs; (c) 2D optical spectra with only MH1 modulation; (d) 2D optical spectra with only MH2 modulation; (e) wavelength gap of the two passbands versus MH currents.

The grating samplings adhere to equation (2). The device comprises 215 complete sampling units distributed throughout its cavity, denoted by $n$=215. The chirp rate $C$ is set at 100 nm/mm, with 1st sampling period $P_{s1}$ of 3.079 μm, aligning the +1st order channel near the wavelength of 1550 nm. The 66.9 nm difference in sampling periods can be achieved with high precision using electron-beam lithography (EBL).

The device fabrication process begins by defining 210-nanometer-high Si ridge waveguides and side-wall gratings using EBL, and inductively coupled plasma (ICP) etching. The 10 nm layer of remaining Si prevents the removal of the underlying SiO$_2$ layer when using a hydrofluoric (HF) acid solution to remove the HSQ mask. Subsequently, employing polymethyl methacrylate (PMMA) photoresist and mask, a second round of EBL exposure and dry etching shapes 110 nm high apodized GCs at the input and output ends of the CSG. Following this, a 300 nm thick layer of SiO$_2$ is deposited on the wafer surface via plasma-enhanced chemical vapor deposition (PECVD) to create the buried waveguide. Surface flattening is achieved by spin-coating a 400-nanometer-thick layer of HSQ and annealing at 180°C. Finally, the MHs part, comprised of Ti (20 nm), Pt (160 nm), and Au (50 nm), is defined by metal evaporation using lift-off technology with PMMA photoresist.

In Fig. 4(a), the experimental setup for device characterization at room temperature comprises a superluminescent diode (SLD) serving as the light source, with a central wavelength of 1551 nm and a 3 dB bandwidth of 30 nm. The single-mode fiber (SMF) is linked to the SLD, and a polarization controller (PC) directs the light onto the input GC surface. The output end utilizes a coated lensed single-mode fiber (LSMF) to couple the output light from the surface of the output GC and mitigate unwanted resonances originating from the Fabry-Pérot (FP) cavity between the fiber and the GC. The input and output SMFs are positioned directly above the GCs at an offset angle of 15° from the vertical direction. The output optical signal from the device is observed using an optical spectrum analyzer (OSA) with a resolution of bandwidth (RBW) of 0.06 nm. The output light wavelengths of the device can be tuned by adjusting the injection currents applied to the contact pads of MHs ($I_{MH}$).

Fig. 4(b) shows the transmission spectrum of the device with no injection currents applied to the MHs. The spectrum reveals dual passbands (Q factor: $1.48 \times 10^4$) within a 4 nm range centered around the 1553.4 nm stopband. Band 1 has a -3 dB bandwidth of 27 GHz and a -10 dB bandwidth of 68 GHz, while band 2 features a -3 dB bandwidth of 28 GHz and the same -10 dB bandwidth of 68 GHz. The frequency separation is 109.2 GHz, and the extinction ratio (ER) is 25.7 dB, closely matching the simulation results and demonstrating a balanced and consistent intensity distribution. The on-chip GC demonstrates a coupling loss of 10.2 dB at the wavelength of 1550 nm. The 3.4 nm redshift of the center wavelength compared to the simulation result may be due to etching the ridge waveguide height to 210 nm instead of 220 nm as in the simulation, which increases the effective index and, consequently, the center wavelength. The relatively lower Q factor compared to the simulation may be due to the limited RBW of OSA and fabrication imperfections, such as sidewall and surface roughness of the ridge waveguide, which introduce additional scattering losses, thereby reducing the Q factor. The effective grating coupling coefficient $\kappa$ for the +1st channel is measured to be 102/cm. Fig. 4(c) illustrates the transmission spectrum when only $I_{MH1}$ is tuned from 0 mA to 35 mA. During the scan of $I_{MH1}$ from 0 to 35 mA, the TO effect of the EPS induces a redshift of band 1. The center wavelength of band 2 experiences a slight redshift of 0.012 nm (<RBW) during the modulation of MH1, allowing for independent tuning of a single band. The separation frequency changes from 109.2 GHz to 37.2 GHz. The consistent positioning of the stopband suggests precise alignment and geometric layout of the MHs, with minimal residual heat leakage. Similarly, as depicted in Fig. 4(d), independent modulation of band 2 can also be achieved by solely tuning $I_{MH2}$. When $I_{MH2}$ is tuned from 0 mA to 35 mA while $I_{MH1}$=0 mA, the separation frequency shifts from 109.2 GHz to 186.1

GHz, while band 1 has a wavelength deviation of only 0.03 nm (<RBW). During this process, band 2 maintains its narrowband characteristics, with the -3 dB bandwidth increasing slightly from 28 GHz to 29.4 GHz. Furthermore, as shown in Fig. 4(e), analysis of the changes in the passband spacing during MH modulation from 0 to 35 mA reveals continuous modulation capability ranging from 37.2 GHz to 186.1 GHz, enabling dynamic tuning of the separation of the passbands. According to the simulation results in Fig. 2(c) and Fig. 2(d), the tuning wavelength gap can range up to 259 GHz if the MH wire is optimized to prevent breakage when a high injection current (>35 mA) is applied. Experiments on synchronous modulation of the dual passbands by tuning $I_{MH1}$ and $I_{MH2}$ from 0 mA to 35 mA simultaneously were also conducted, maintaining a relatively stable wavelength spacing and further confirming the independent tuning characteristic of the dual passbands.

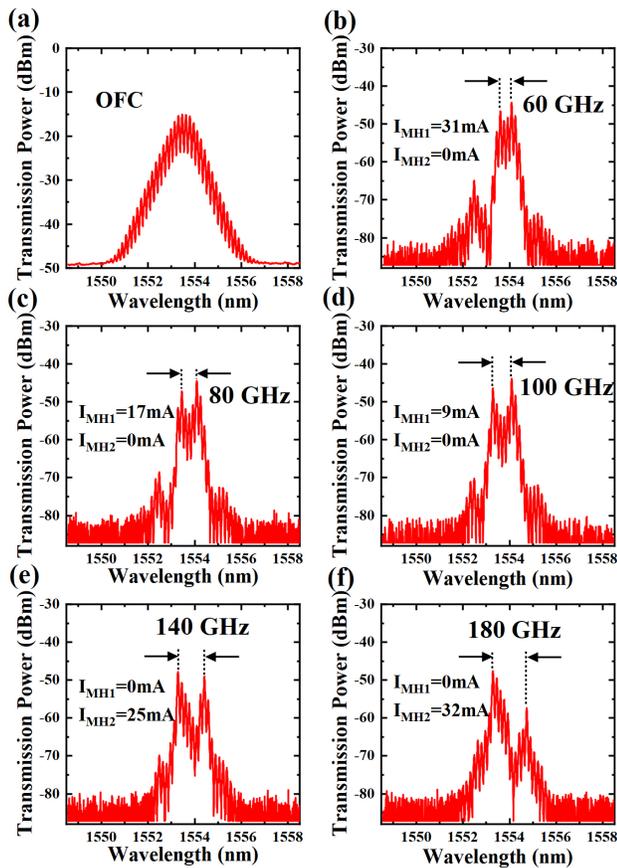

**Fig. 5.** Optical spectra of (a) unfiltered optical frequency comb, and with various filter settings with frequency separations of (b) 60 GHz, (c) 80 GHz, (d) 100 GHz, (e) 140 GHz, (f) 180 GHz.

We also employed an ultrafast optical clock (UOC) driven optical frequency comb (OFC) with a central wavelength of 1553.4 nm and a frequency interval of 20 GHz (Fig. 5(a)) to characterize our device. The output spectra shown in Fig. 5(b) to 5(f) were obtained by modulating the driving currents of MH1 and MH2, as depicted in the respective figures. An analysis of the optical frequency division (OFD) characteristics of the dual-wavelength PF revealed that the output spectra aligned with the trends observed using an SLD source shown in Fig. 4(e). Specifically, when only MH1 was modulated from low current to high current, band 1 initially approached band 2, decreasing the separation frequency of the PF. Conversely, when only MH2 was modulated, band 2 moved away from band 1, increasing the separation frequency of the PF. The figures illustrate five distinct filtering configurations with separations of 60 GHz, 80 GHz, 100 GHz, 140 GHz, and 180 GHz, achieving a tunable filtering range from 60 GHz to 180 GHz, with the corresponding $I_{MH1}$ and $I_{MH2}$. The 180 GHz configuration represents the maximum OFD spacing, limited by the 20 GHz OFC interval and the 27 GHz -3 dB channel bandwidth.

In summary, we present a dual-band independently tunable PF integrated on an SOI platform. Employing an SLD as the input source, the filter exhibits continuous frequency tuning capabilities spanning from 37.2 GHz to 186.1 GHz. Further characterization is conducted using an OFC with a 20 GHz interval, showcasing filtering capabilities from 60 GHz to 180 GHz. This dual-band tunable PF holds promise for applications in wireless communication, radar, and optical communication systems. Moreover, our technology can be extended for Terahertz (THz) photonic filter applications.

**Funding.** This work was supported by the U.K. Engineering and Physical Sciences Research Council (EP/R042578/1).

**Acknowledgments.** We would like to acknowledge the staff of the James Watt Nanofabrication Centre at the University of Glasgow for their help in fabricating the devices.

**Disclosures.** The authors declare no conflict of interest.

**Data availability.** Data underlying the results presented in this paper are not publicly available at this time but may be obtained from the authors upon reasonable request.